\documentclass[aps,prb,twocolumn,groupedaddress]{revtex4-1}

\usepackage{graphicx}
\usepackage{dcolumn}
\usepackage{bm}
\usepackage{amsmath}
\usepackage{amssymb}
\usepackage{latexsym}
\usepackage{epsfig}
\usepackage{amsbsy}
\usepackage{array}
\usepackage{amssymb}
\usepackage{setspace}
\usepackage{bbm}
\usepackage{color}

\usepackage{tabularx}
\usepackage{ragged2e}
\usepackage{booktabs}

\usepackage{setspace}

\def\sint{\ifmmode{- \!\!\!\!\!\! \int}
    \else{\hbox{$- \!\!\!\! \int \ $}}\fi}

\begin{document}

\bibliographystyle{apsrev4-1}

\title{Theory for Spin Selective Andreev Reflection in Vortex Core of Topological Superconductor: Majorana Zero Modes on Spherical Surface and Application to Spin Polarized Scanning Tunneling Microscope Probe}

\author{Lun-Hui Hu$^{1,2}$}
\author{Chuang Li$^{1,2}$}
\author{Dong-Hui Xu$^{3}$}
\author{Yi Zhou$^{1,2}$}
\email{yizhou@zju.edu.cn}

\author{Fu-Chun Zhang$^{1,2}$}
\email{fuchun@hku.hk}

\affiliation{$^1$Department of Physics, Zhejiang University, Hangzhou, Zhejiang, 310027, China}
\affiliation{$^2$Collaborative Innovation Center of Advanced Microstructures, Nanjing 210093, China}
\affiliation{$^3$Department of Physics, Hong Kong University of Science and Technology, Clear Water Bay, Hong Kong, China}

\date{\today}

\begin{abstract}
  Majorana zero modes (MZMs) have been predicted to exist in the topological insulator (TI)/superconductor (SC) heterostructure.  Recent spin polarized scanning tunneling microscope (STM) experiment \cite{arxiv_mzm_our_2016} has observed spin-polarization dependence of the zero bias differential tunneling conductance at the center of vortex core, which may be attributed to the spin selective Andreev reflection, a novel property of the MZMs theoretically predicted in 1-dimensional nanowire.\cite{he_prl_2014} Here we consider a helical electron system described by a Rashba spin orbit coupling Hamiltonian on a spherical surface with a s-wave superconducting pairing due to proximity effect.  We examine in-gap excitations of a pair of vortices with one at the north pole and the other at the south pole. While the MZM is not a spin eigenstate, the spin wavefunction of the MZM at the center of the vortex core, $r=0$, is parallel to the magnetic field, and the local Andreev reflection of the MZM is spin selective, namely occurs only when the STM tip has the spin polarization parallel to the magnetic field, similar to the case in 1-dimensional nanowire\cite{he_prl_2014}. The total local differential tunneling conductance consists of the normal term proportional to the local density of states and an additional term arising from the Andreev reflection.  We also discuss the finite size effect, for which the MZM at the north pole is hybridized with the MZM at the south pole. We apply our theory to examine the recently reported spin-polarized STM experiments and show good agreement with the experiments.
\end{abstract}


\maketitle

\section{Introduction}
In condensed matter physics, Majorana\cite{Majorana_2008} zero modes (MZMs) are a special type of Bogoliubov quasiparticle excitations with non-Abelian statistics,
which have been proposed to be building blocks for quantum information and quantum computation\cite{kitaev_pu_2001,kitaev_aop_2003}.
There have been a number of theoretical proposals\cite{nayak_rmp_2008,elliott_rmp_2015} to realize MZMs in condensed matter systems,
such as $\nu=5/2$ fractional quantum hall system\cite{more_read_1991}, chiral p-wave superconductor\cite{sarma_prb_2006},
topological insulator(TI)/s-wave superconductor(SC) interfaces with MZM in the vortex core\cite{fu_prl_2008},
proximity-induced superconductor for spin-orbit coupled nanowires\cite{oreg_prl_2010,lutchyn_prl_2010} and spin-orbit coupled semiconductor\cite{sau_prl_2010,tewari_aop_2010,alicea_prb_2010} with externally applied Zeeman field,
and ferromagnetic atoms in proximity to superconductors\cite{choy_prb_2011,nadj_perge_prb_2013}.
There also exist various experimental evidences for MZMs in these proposed systems\cite{Wilczek_np_2009,mourik_science_2012,deng_nano_lett_2012,das_np_2012,williams_np_2012, xu_prl_2014,perge_science_2014, xu_prl_2015,kjaergaard_arxiv_2016}.

Very recently, He \emph{et.al.}\cite{he_prl_2014} have predicted that a MZM at the end of a nanowire may induce spin selective Andreev reflection (SSAR).
An electron with the same spin of the MZM will undergo an Andreev reflection, while an electron with opposite spin does not.  This SSAR is a novel property of the MZMs and is different from the ordinary Andreev reflection(AR)\cite{BTK_physics_1982,kashiwaya_rep_pro_2000, tinkham_book_2004}.
This property may allow us to reveal spin degrees of freedom of the MZMs.  However, in 1D nanowire systems, it always requires a large Zeeman term to host MZMs,  which may make it difficult to attribute the spin polarization dependence to the MZMs.

Fu and Kane \cite{fu_prl_2008} proposed that MZM is localized inside the vortex core in TI/SC heterostructure, and they showed this explicitly by solving Bogliubov de Gennes equations (BdG)\cite{desuperconductivity_1999}.
Experimentally, the MZMs in such a system have been demonstrated by STM based on zero bias peak (ZBP) in the heterostructure Bi$_2$Te$_3$/NbSe$_2$, made of TI thin film Bi$_2$Te$_3$ on the top of SC NbSe$_2$ \cite{xu_prl_2014,xu_prl_2015}.
Most recently, strong new evidence for the MZM inside the vortex core is reported by using spin-polarized STM\cite{arxiv_mzm_our_2016}. The experiment has clearly shown spin polarization dependence of the differential tunneling conductance $dI/dV(E,r=0)$.
In this paper, we present a systematic model calculation to examine the SSAR inside the vortex core of TI/SC.

We consider a helical metal described by a Rashba spin-orbit coupling Hamiltonian on a spherical surface of radius $R$.  Superconductivity is introduced by proximity effect, and electronic structure in a vortex state is studied.
At the center of the vortex core, $r=0$, the spin component of the MZM is parallel to the magnetic field, and the local Andreev reflection of the MZM is spin selective, and only occurs when the STM tip has the spin polarization parallel to the magnetic field.
The first quasiparticle state has the same amplitude of orbital wavefunction with, but opposite spin polarization to the MZM at $r=0$.
This leads to the approximately same local density of states and the normal differential tunneling conductance for the spin parallel and anti-parallel to the magnetic field.
We also discuss the finite size effect. We apply our theory to examine the recently reported spin-polarized STM experiments and show good agreement with the experiments.

The paper is organized as follows. In Sec.\ref{sec-1-models-methods}, the BdG equation is introduced and we adopt spherical geometry to solve the eigenfunction problem.
The numerical results are presented systematically in Sec.\ref{sec-2-solve-bdg-equ}.  Method and results of transport calculation for Andreev reflection(AR) at the center of the vortex core is discussed in detail in Sec.\ref{sec-3-AR}.
A summary and conclusion are presented in Sec.\ref{sec-4-conclusion}. In the Appendix.\ref{sec-3B-BTK-theory}, the vortex-free BdG equation will be addressed and the well known Blonder-Tinkham-Klapwijk (BTK) theory is reproduced by using our method.

\section{Vortex states in proximity-induced topological superconductor on spherical surface}\label{sec-1-models-methods}
In this section we discuss vortex states in a topological superconductor, which is modelled by a helical metal with proximity-induced superconductivity.
The helical metal represents the surface states of a 3D topological insulator. We consider spherical geometry, in which electrons in the helical metal are confined on a spherical surface.  In this geometry, the boundary is closed.
We will start with a non-interacting electron system with a Rashba spin-orbit coupling, then discuss such a helical metal under the proximity effect of superconductivity in the absence of vortices. Finally, we will also discuss the in-gap vortex states of the system.
\subsection{2D Helical metal on spherical surface}
The surface state of a 3D topological insulator may be described by non-interacting helical electrons on the x-y plane. The single electron Hamiltonian reads,
  \begin{align}
     \mathcal{H}_0 &=\alpha(\sigma\times\vec{p})\cdot\vec{z}-\mu,     \label{eq:ham-surface}
  \end{align}
where $\alpha$ is the spin-orbit coupling strength, which will be assumed to be positive throughout this paper without loss of generality.
$\vec {\sigma}$ is made of the three Pauli matrices, $\vec p$ is the momentum confined in the x-y plane, and $\mu$ is the chemical potential.
The Hamiltonian in the x-y plane can be generalized to spherical surface of radius R, by using the expression $\vec{p}=-i\left\lbrack \vec{\nabla}-\hat{R}\left(\hat{R}\cdot\vec{\nabla}\right)\right\rbrack$,
where $\hat{R}=\vec{R}/R$ and we set $\hbar=1$. The Hamiltonian for a helical metal on a spherical surface is then given by

\begin{align}
     \mathcal{H}_0 &=-\frac{\alpha}{R}\vec{L}\cdot\vec{\sigma} -\mu,  \label{eq:ham-surface-2}
\end{align}
where $\vec L$ is the orbital angular momentum. The single particle Hamiltonian Eq.\eqref{eq:ham-surface-2} can be solved easily.
Define the total angular momentum $\vec J =\vec L +\vec S$.  $H_0$ commutes with $\vec J$, the z-component of $\vec{J}$ ($J_z$), $\vec L$, and electron spin operator $\vec{S}=\vec{\sigma}/2$.
The total angular momentum eigenvalue $j=1/2$ if $l=0$, and $j = l \pm 1/2$ if $l\neq 0$. The eigen-energy $E_0$ of Eq.\eqref{eq:ham-surface-2} is given by
\begin{align}\label{eq-energy-h0-1}
\begin{cases}
j=l+1/2: \quad E^0_{+}(l)= -2\alpha \frac{l}{R}-\mu,  \\ \\
j=l-1/2: \quad E^0_{-}(l)= 2\alpha \frac{l+1}{R}-\mu.
\end{cases}
\end{align}
and the 2-component eigen-states in a compact form,
\begin{align}\label{eq-2-component-wave-func-vortex-free}
  &\vert j=l\pm 1/2,j_z,l,s=1/2\rangle \\
  =\;& \alpha_{\pm}\vert l,j_z-1/2\rangle \otimes\vert\uparrow\rangle
   + \beta_{\pm}\vert l,j_z+1/2\rangle\otimes\vert \downarrow\rangle, \nonumber
\end{align}
where the Clebsch$-$Gordan(CG) coefficients are $\alpha_{\pm}=\pm\sqrt{(l\pm j_z+1/2)/(2l+1)}=\pm\beta_{\mp}$. There are $2j+1$ degenerate states for a given eigen-energy, corresponding to different eigenvalue of $J_z$.
Note that the states in $j=l+1/2$ branch have negative energies without a lower energy bound, similar to the case in planar geometry. This will, however, not affect the basic physics near the Fermi energy which we wish to address in this paper.

\subsection{Proximity-induced superconducting state in a helical metal}
We now consider the helical metal Hamiltonian with an additional pairing term describing proximity induced superconductivity. We shall first consider a uniform pairing case that is free of vortices, described by
\begin{align}\label{eq-ham-normal-sc}
    \mathcal{H}_{\Delta} &=  \Delta_0 \sigma_0\otimes\tau_x,
\end{align}
where we use standard Nambu representation, for the field operator $\hat{\Psi}(\vec{r})$,
\begin{align}
  \hat{\Psi}(\mathbf{r})=\lbrack \hat{c}_\uparrow(\mathbf{r}),\hat{c}_\downarrow(\mathbf{r}), \hat{c}_\downarrow^\dagger(\mathbf{r}), -\hat{c}_\uparrow^\dagger(\mathbf{r}) \rbrack ^{T},
\end{align}
In Eq.\eqref{eq-ham-normal-sc}, we have assumed the proximity induced superconducting order parameter $\Delta$ to be independent of the azimuthal angle $\theta$ and the polar angle $\phi$.  The total Hamiltonian then reads,
\begin{align}\label{eq-bdg-ho-hdelta-vortex-free-1}
  \mathcal{H} &= \left(
                      \begin{array}{cc}
                        \mathcal{H}_0 & \Delta_0 I \\
                        \Delta_0 I & -\sigma_y\mathcal{H}_0^\ast\sigma_y \\
                      \end{array}
                    \right)  \nonumber \\
  &= \left(-\frac{\alpha}{R}\vec{L}\cdot\vec{\sigma} -\mu\right)\otimes\tau_z + \Delta_0 \sigma_0\otimes\tau_x,
\end{align}
where $\vec{\tau}$ are Pauli matrices, with the two components representing particle-hole degrees of freedom, while  $\vec{\sigma}$ representing spin degree of freedom,
and $\sigma_0$ the identity matrix. Note that the total angular momentum $\vec{J}=\vec{L}+\vec{S}$ still commutes with $\mathcal{H}$ in Eq.\eqref{eq-bdg-ho-hdelta-vortex-free-1}, the eigen-energies are given by
\begin{align}\label{eq-solution-vortex-free-energy}
\begin{cases}
  j=l+1/2:\quad E_{\pm}(l) = \pm \sqrt{\left\lbrack E^0_{+} (l)\right\rbrack^2+\Delta_0^2}, \\ \\
  j=l-1/2:\quad E_{\pm}(l) = \pm \sqrt{\left\lbrack E^0_{-} (l)\right\rbrack^2+\Delta_0^2}.
\end{cases}
\end{align}
where the corresponding 4-component eigen-function (see Eq.\eqref{eq-2-component-wave-func-vortex-free}) is $A\vert j=l\pm1/2,j_z,l,s=1/2\rangle \oplus B\vert j=l\pm1/2,j_z,l,s=1/2\rangle$ formally, in which the coefficients $A$ and $B$ satisfies
\begin{align}
  A\Delta_0 = B\left\lbrack E_{\pm}^{0}\pm\sqrt{\left(E_{\pm}^{0}\right)^2+\Delta_0^2}\right\rbrack,
\end{align}
where the normalization condition $A^2+B^2=1$ gives $A=f/\sqrt{f^2+\Delta_0^2}$ and $B=\Delta_0/\sqrt{f^2+\Delta_0^2}$, with $f=E_{\pm}^{0}+\sqrt{\left(E_{\pm}^{0}\right)^2+\Delta_0^2}\text{ or } E_{\pm}^{0}-\sqrt{\left(E_{\pm}^{0}\right)^2+\Delta_0^2}$.

\subsection{Vortex states in proximity-induced superconducting state of helical metal}
\begin{figure}[!htbp]
  \centering
  \includegraphics[width=1.8in]{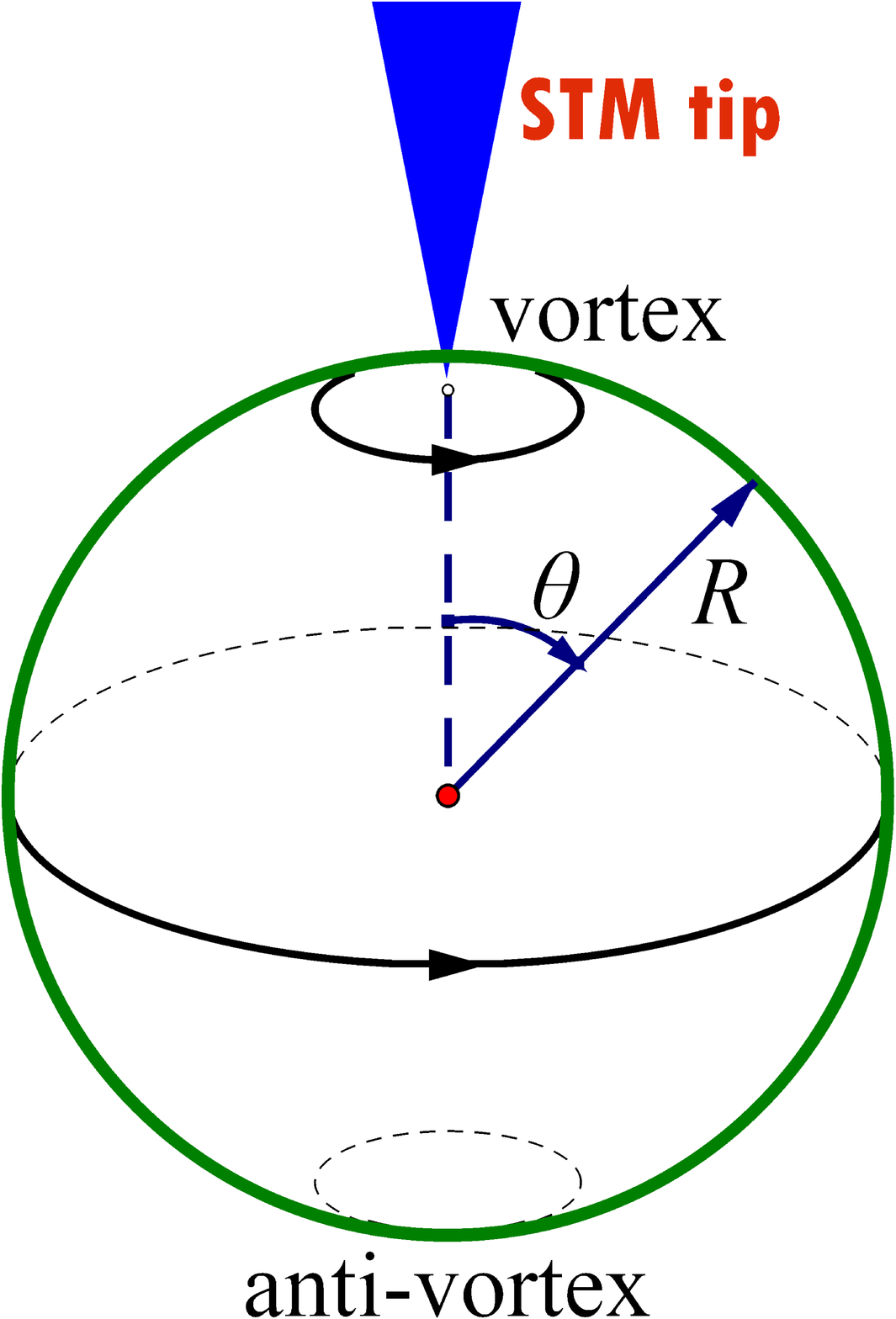}
  \caption{\label{fig-add-1-sketch-sphere} Sphere geometry for vortex problem. There exists a pair of vortex and anti-vortex. The vortex locates at the north pole and the anti-vortex locates at the south pole.}
\end{figure}

We proceed to consider the vortex case.  The pairing Hamiltonian takes the form
\begin{align}
    \label{eq-ham-vortex-first}
    \mathcal{H}_{\text{BdG}}(\Delta) &=  \left\lbrack \Delta(\theta) e^{i\phi}\right\rbrack \sigma_0.
\end{align}
In Eq.\eqref{eq-ham-vortex-first}, the factor $e^{i\phi}$ describes a vortex with the winding number $n=1$. Then we will use a numerical method to solve the vortex problem.
The whole Bogoliubov de-Genes Hamiltonian in the presence of the vortex consists of Eq.\eqref{eq:ham-surface-2} and Eq.\eqref{eq-ham-vortex-first},
  \begin{align}\label{eq:bdg-ham-ti-sc}
     \mathcal{H}_{\text{BdG}}  &= \left(
                      \begin{array}{cc}
                        \mathcal{H}_0 & \Delta(\theta)e^{i\phi} \\
                        \Delta(\theta)e^{-i\phi} & -\sigma_y\mathcal{H}_0^\ast\sigma_y \\
                      \end{array}
                    \right)  \nonumber \\
                      &=  \left(-\frac{\alpha}{R}\vec{L}\cdot\vec{\sigma} -\mu\right)\otimes\tau_z + \Delta(\theta) e^{i\phi}I\otimes\tau_x.
  \end{align}
We note that there is a vortex-antivortex pair, one of which (say, the vortex) locates at the north pole of the sphere and the other one (the anti-vortex) is at the south pole\cite{kraus_prl_2008,kraus_prb_2009}, see Fig.\ref{fig-add-1-sketch-sphere}.
In this paper, we assume the gap function $\Delta(\theta)=\Delta_0\tanh\left\{R\sin(\theta)/\xi_0\right\}$, where $\xi_0$ characterizes the size of the vortex core.

In our model, we neglect the small Zeeman term and the vector potential. The neglect of the vector potential will be a good approximation to describe low magnetic field case, such as the experiment in Ref.\cite{arxiv_mzm_our_2016},
where the external magnetic B is lower than 0.1 Tesla. This approximation greatly simplifies the matter and is expected not to change our result qualitatively. Define Bogoliubov quasi-particle operators as (using $r=R\theta$)
  \begin{align}
     \hat{\gamma}^\dagger = \int d\mathbf{r} \left\lbrack \sum_{\sigma} u_{\sigma}(\mathbf{r})\hat{c}_\sigma^\dagger(\mathbf{r}) + v_\sigma(\mathbf{r})\hat{c}_\sigma(\mathbf{r}) \right\rbrack.
  \end{align}
In passing, we remark that the necessary condition for the existence of the MZM is \(\hat{\gamma}^\dagger = \pm\hat{\gamma}\).  The spectra of the excitations of this system can be found by solving the eigenvalues problem in 2D coordinate $\{\theta,\phi\}$,
similar to the calculations reported in Ref.\cite{sau_prb_2010}. Let $\vert\Phi_m\rangle$ be the 4-component wavefunction of the field $\hat{\Psi}(\vec{r})$, and the eigenvalue problem is given by
\begin{align}\label{eq:eigenvalues-problem}
    \mathcal{H}_{\text{BdG}}\vert\Phi_m\rangle = E_m \vert\Phi_m\rangle,
\end{align}
with
\begin{align}
\label{eq-psi-m-wave-function}
\vert\Phi_m\rangle = \left( e^{im\phi}u_1,e^{i(m+1)\phi}u_2, e^{i(m-1)\phi}v_1, e^{im\phi}v_2 \right)^{T},
\end{align}
where $m$ is the eigenvalue of $K_z$ that will defined in Eq.\eqref{eq-total-angular-kz} below, and $u_1, \, u_2, \, v_1, \, v_2$ are real functions of $\theta$ and $m$, but are independent of $\phi$.
As we shall see below, there will be a pair of MZMs\cite{sau_prl_2010} in the channel $m=0$. And the $m=\pm 1$ channel gives the first quasi-particle excitation\cite{sau_prb_2010,sau_prb_2010-2}.
Note that the total angular momentum $J_z=L_z+\sigma_z/2$ does not commute with the Hamiltonian in Eq.\eqref{eq:bdg-ham-ti-sc}, because of the winding phase factor $e^{i\phi}$ in the gap function.
To solve the BdG Hamiltonian numerically, we observe that this Hamiltonian has a combined spin-orbit-pseudo-spin (pseudo-spin here refers to the particle-hole degree of freedom) rotational symmetry along the z-axis.
This symmetry can be expressed compactly by noting that the Hamiltonian in Eq.\eqref{eq:bdg-ham-ti-sc} commutes with a generalized total angular momentum including the pseudo-spin $\tau_z$, so we have,
\begin{align}\label{eq-total-angular-kz}
      K_z = L_z + \frac{1}{2}\left(\sigma_z-\tau_z\right)  \quad \Rightarrow \quad \left\lbrack K_z,\mathcal{H}_{\text{BdG}} \right\rbrack=0.
\end{align}
The BdG Hamiltonian can be decomposed into block-diagonal form, with each block corresponding to a generalized total angular momentum \(m\) (quantum number of \(K_z\)), namely,
\begin{subequations}
\begin{align}
      \label{eq:two_good_quamtum_operator_h_jz}
      \mathcal{H} \vert\Phi_m \rangle &= E \vert \Phi_m \rangle, \\
       \label{eq:two_good_quamtum_operator_h_jz-2}
       K_z  \vert\Phi_m \rangle &= m \vert\Phi_m\rangle,
\end{align}
\end{subequations}
where $\vert\Phi_m\rangle$ is given in Eq.\eqref{eq-psi-m-wave-function}. The four-component eigenfunction/basis in $\vert\Phi_m\rangle$ may be expressed in terms of the spherical harmonic functions,
\begin{subequations}\label{equv12}
\begin{align}
       &e^{im\phi}\;u_1(m)   = \sum_{l}a_l Y_l^m  \label{eq-expansion-u1-u2-v1-v2-to-Ylm-1},\\
       &e^{i(m+1)\phi}\;u_2(m+1) = \sum_{l} b_l Y_l^{m+1} \label{eq-expansion-u1-u2-v1-v2-to-Ylm-2},\\
       &e^{i(m-1)\phi}\;v_1(m-1) = \sum_{l} c_l Y_l^{m-1} \label{eq-expansion-u1-u2-v1-v2-to-Ylm-3},\\
       &e^{im\phi}\;v_2(m) = \sum_{l} d_l Y_l^{m}\label{eq-expansion-u1-u2-v1-v2-to-Ylm-4},
\end{align}
\end{subequations}
with \(Y_l^m(\theta,\phi) = P_l^m(\cos\theta) e^{im\phi}/\sqrt{2\pi}\), and $P_l^m$ the associated Legendre polynomial. The above eigen-state problem can be solved numerically.
The particle-hole symmetry is reflected as below. If we transform $m\to-m$, then we have $E_m\to-E_m$ and $\{u_1,u_2,v_1,v_2\}\to \{-v_2,v_1,u_2,-u_1\}$. Note that these wavefunctions are all real.

The system is invariant under rotation $\hat{\mathcal{O}}=\hat{P}\sigma_z\otimes\tau_0$, which communicates with the Hamiltonian,
\begin{align}
  \left\lbrack \hat{\mathcal{O}},\mathcal{H}_{\text{BdG}}\right\rbrack=0.
\end{align}
It means that the total Hamiltonian remains unchanged by transformation simultaneously both in real space $\hat{P}:\theta\to(\pi-\theta)$ and in spin subspace $\sigma_z: \sigma_x\to-\sigma_x, \sigma_y\to-\sigma_y$.
This symmetry is important to analyze the two-fold degeneracy of in-gap quasi-particle states in the large radius limit. Assume $\mathcal{H}_{\text{BdG}}\vert \Phi_m^{\pm}\rangle = E_m^{\pm}\vert \Phi_m^{\pm}\rangle$ and $\hat{\mathcal{O}}\vert \Phi_m^{\pm}\rangle = \pm\vert \Phi_m^{\pm}\rangle$,
we can write down the in-gap quasi-particle wavefunctions in the form,
\begin{widetext}
\begin{align}\label{eq-degenracy-state}
  \vert\Phi_m^{\pm}\rangle = e^{im\phi}\left( \lbrack u_1(\theta) \pm u_1(\pi-\theta)\rbrack, \, e^{i\phi}\lbrack u_2(\theta) \mp u_2{\pi-\theta}\rbrack, \, e^{-i\phi}\lbrack v_1(\theta)\pm v_1(\pi-\theta)\rbrack, \, \lbrack v_2(\theta)-v_2(\pi-\theta) \rbrack \right)^{T}.
\end{align}
\end{widetext}
To see the physical interpretation, we can define $\left\vert \Phi_m^{\text{N}}\right\rangle = \left(\vert\Phi_m^{+}\rangle + \vert\Phi_m^{-}\right)/\sqrt{2}$,
which is localized at the north pole and vanishes at the south pole.
Meanwhile, we also have $\left\vert\Phi_m^{\text{S}}\right\rangle = \left(\vert\Phi_m^{+}\rangle - \vert\Phi_m^{-}\right)\rangle /\sqrt{2}$, localized in the south pole.
Therefore, the symmetry $\hat{\mathcal{O}}$ gives us the degenerated states $\left(\left\vert \Phi_m^{\text{N/S}}\right\rangle\right)$ in the energy spectrum in the large radius limit.
However, this degeneracy will be lifted a little due to the hybridization between these two states in our numerical simulation, which is related to the finite size effect(finite radius $R$) and will be discussed in Sec.\ref{sec-2B-finite-size-effect} in details.

Lastly, we also wish to emphasize that once all the eigen-energies and eigen-wavefunctions are obtained, the Green's function thereby the transport properties can be calculated, as we will discuss in Sec.\ref{sec-3-AR}.

\section{Numerical Results}\label{sec-2-solve-bdg-equ}
In this section, we present the solutions of the BdG equations for the proximity-induced vortex states in topological superconductor on a spherical surface and discuss the results in connection with recent experiments.

\subsection{Energy spectra and wavefunctions}\label{sec-2A-dispersion-wave}
In our numerical calculations, we use parameters to approximately model the experiment in Ref~\cite{xu_prl_2015}, in which the MZMs in Bi\(_2\)Se\(_3\)/NbSe\(_2\) heterostructure has been detected.
We set the parameters as follows, the coherence length $\xi_0=35$ nm is chosen as the length unit, the radius of the sphere $R=50\xi_0$, the superconducting gap far away from the vortex core $\Delta_0=1$ meV,
the chemical potential $\mu=90$ meV and the spin-orbit coupling $\alpha=30\xi_0$ meV$\cdot$nm.
Note that the coherence length, the chemical potential, and the gap function are comparable to the extracted experiment data in  Bi\(_2\)Se\(_3\) of $35$ nm, $100$ meV, and $1$ meV, respectively\cite{xu_prl_2015}.
The Fermi velocity $v_F=\alpha \xi_0$=1.05 nm\(\cdot\)eV in our simulation is about 4 times larger than the experiment data of 0.27 nm\(\cdot\)eV\cite{Chen_science_2009,zhang_nat_phy_2009}.
The choice of a large spin-orbit coupling or a large Fermi velocity in our simulation is for the technique reasons in the calculations to avoid handling spherical harmonic functions of very large $l$ ,
which turns to be quite challenging. The parameters used in our model calculations are close to those values used in the reported simulation in Ref\cite{arxiv_mzm_our_2016}.
In our numerical calculations, we take a cutoff in the orbital angular momentum $l$ around $200$, which is sufficient to get the precise low energy eigenvalues and the accurate corresponding spatial wavefunctions.

\begin{figure}[!htbp]
  \centering
  \includegraphics[width=3.3in]{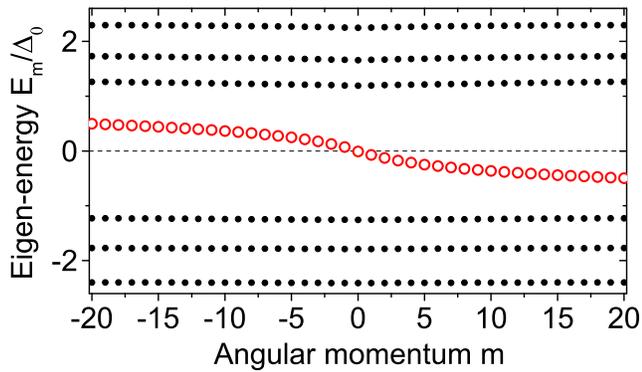}
  \caption{\label{Fig-1-ADD-A-dispersion-vortex-and-free} Energy spectra for the vortex problem. $E_m$ is plotted as a function of angular momentum $m$ (quantum number of $K_z$ in Eq.\eqref{eq-total-angular-kz}).
  Red hollow circular({\color{red}$\circ$}) represents localized in-gap bound states, and black solid circular($\bullet$) represents bulk states.
  The energy discretization for bulk states is due to finite size effect.
  Here we choose parameters as $\Delta_0=1$ meV, $\xi_0=35$ nm, $R=50\xi_0$, $\alpha=30$ meV and $\mu=90$ meV.}
\end{figure}

The energy spectra of the model Hamiltonian Eq.\eqref{eq:bdg-ham-ti-sc} are plotted in Fig.~\ref{Fig-1-ADD-A-dispersion-vortex-and-free}, where red circles represent several localized in-gap states in the vortex core.
As comparison, we plot the energy spectra of vortex-free case in Fig.\ref{fig-add-dispersion-vortex-free} in Appendix \ref{sec-3B-BTK-theory}.
The energy for MZM is about $E_0=10^{-4}$ meV due to the finite size effect.

\begin{figure}[!htbp]
  \centering
  \includegraphics[width=3.1in]{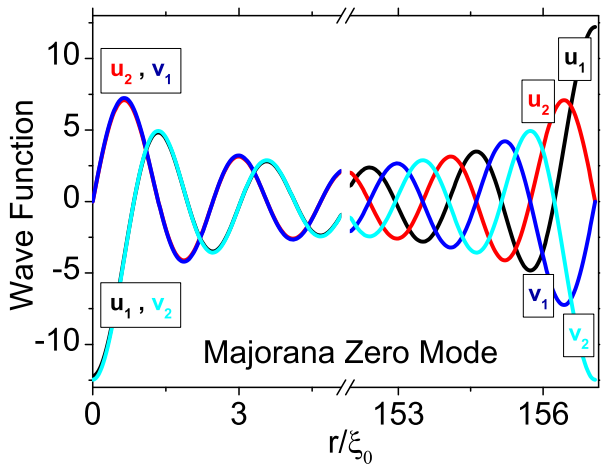}
  \caption{\label{Fig-1-ADD-B-wave-func-MZM} Wavefunction for Majorana zero mode $E_{0}^{-}\approx -10^{-4}$ meV, where $u_1(\theta)=-u_1(\pi-\theta)$, $u_2(\theta)=u_2(\pi-\theta)$, $v_1(\theta)=-v_1(\pi-\theta)$ and $v_2(\theta)=v_2(\pi-\theta)$.}
\end{figure}

The wavefunction for MZM is plotted in Fig.~\ref{Fig-1-ADD-B-wave-func-MZM}. We can see that around north pole: $u_1=v_2$ and $u_2=v_1$, so that $\gamma=\gamma^\dagger$ (necessary condition for MZM).
It worth noting that the spin of MZM is fully polarized at the vortex core center and parallel to the magnetic field (spin-up), say, $u_1=v_2\neq 0$ and $u_2=v_1=0$ at $r=0$.

\begin{figure}[!htbp]
  \centering
  \includegraphics[width=3.1in]{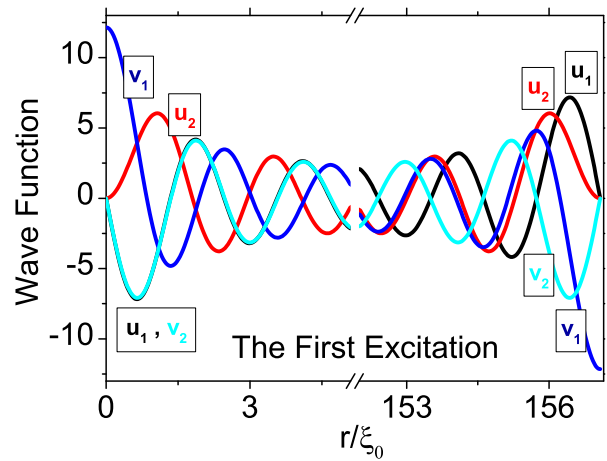}
  \caption{\label{Fig-1-ADD-C-wave-func-the-first-excited-s} Wavefunction for the first excited state $E_1^{-}\approx -0.05$ meV, where $u_1(\theta)=-u_1(\pi-\theta)$, $u_2(\theta)=u_2(\pi-\theta)$, $v_1(\theta)=-v_1(\pi-\theta)$ and $v_2(\theta)=v_2(\pi-\theta)$.}
\end{figure}

Fig.~\ref{Fig-1-ADD-C-wave-func-the-first-excited-s} shows the wavefunction of the first excited state $E_1^{-}$ (quasi-hole).
It is clear that the spin is still fully polarized at the vortex core center but opposite to the magnetic field (spin-down).
Note that there is only one nonzero component in the wavefunction at the vortex core center, namely, $v_1(r=0)\neq0$ and $u_1(r=0)=u_2(r=0)=v_2(r=0)=0$.
Thus this state will not contribute to the anomalous part of Green's function, in Eq.\eqref{eq-green-function-definition}.
Therefore, it will never contribute to Andreev reflection but will only contribute to local density of state, which will be discussed in details in Sec.\ref{sec-3-AR}.
The energy of the first excited state in Fig.\ref{Fig-1-ADD-C-wave-func-the-first-excited-s} is found to be $E_1\approx0.05$ meV.
The discussion for the ratio between $E_1$ and $\eta$ (smearing factor or STM energy resolution) could be found later in Sec.\ref{sec-2C-LDOS}.

\begin{figure*}[!htbp]
  \centering
  \includegraphics[width=6.0in]{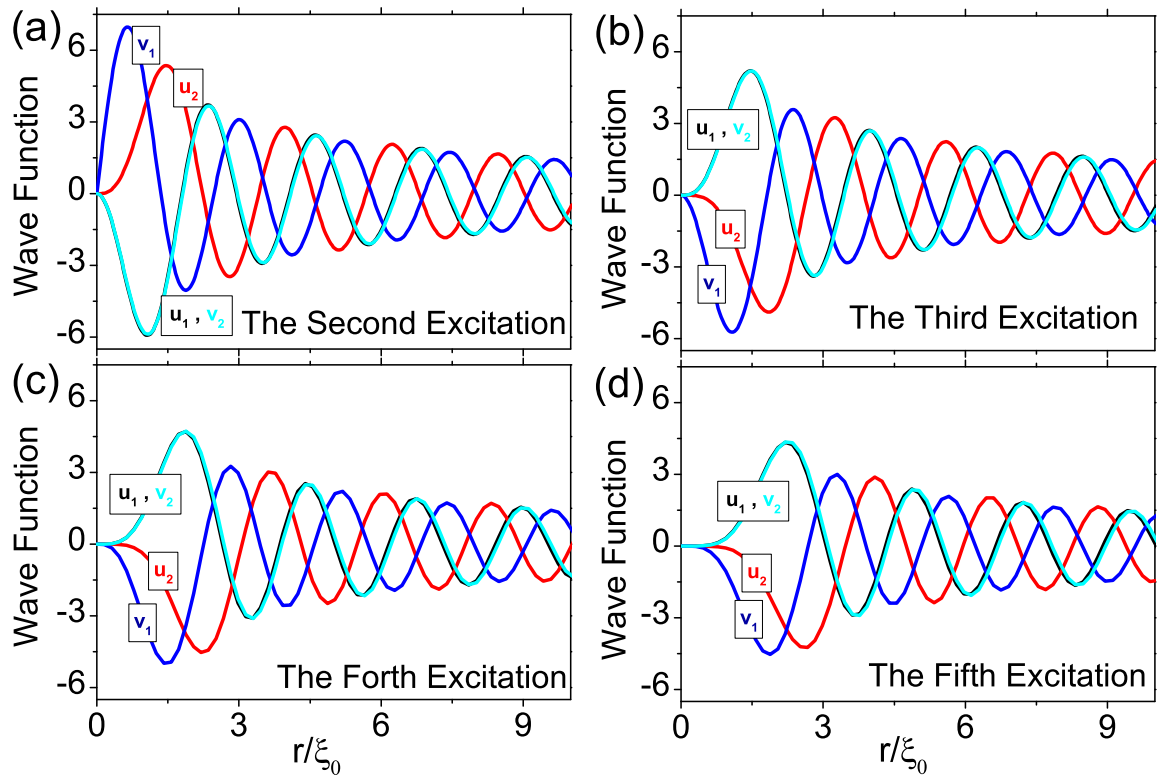}
  \caption{\label{Fig-1-ADD-D-wave-func-high-first-excited-s} Wavefunction for the second (a), third (b), fourth (c), and fifth (d) excited state. }
\end{figure*}

The wavefunctions of the other high angular momentum with $m=2,3,4,5$ are shown in Fig.~\ref{Fig-1-ADD-D-wave-func-high-first-excited-s}.
It is interesting to note that the spherical harmonic function $Y_l^m(\theta,\phi)=0$ at $\theta=0$ and $\theta=\pi$ for all angular momentum $m$ expect $m=0$ channel.
Therefore, all the eigen-wavefunctions $\vert\Phi_m\rangle$ with $\vert m\vert >1$ in Eqs.~\eqref{equv12} for the in-gap states in the vortex core have zero amplitudes at $\theta=0$ and $\theta=\pi$.
This property will greatly simplify the calculations of the local tunneling conductance at the vortex center discussed in Sec.\ref{sec-2C-LDOS}.
The energy for the second excited state is $E_2\approx0.1$ meV.
Finally, we would like to point out that the low lying energy separation of the quasi-particle states in our present calculations appears to be larger than the estimated one in a vortex core which is given by $\Delta_0^2/\mu \approx 0.01$ meV in Ref.\cite{xu_prl_2015},
which could be due to the discrete energy spectra in the free electron system of the model.

\subsection{Two-vortex hybridization (Finite size effect)}\label{sec-2B-finite-size-effect}
In this subsection, we shall study the two-vortex-hybridization problem through the finite size effect in our model.
For experiments, increasing external magnetic field will shorten the distance between two nearest-neighbour vortices on the Abrikosov lattice.
So that two-vortex-hybridization effect will be significant in the presence of higher magnetic field.
Theoretically, we have a pair of vortex and anti-vortex located at north and south pole on the sphere respectively.
Decreasing the sphere radius while keeping coherence length unchanged will give rise to stronger hybridization between the vortex and the anti-vortex and lift the two-fold degeneracy, and vice versa.
Therefore, the two-vortex problem can be reflected as finite size effect in our model.

To study the finite size effect, let us take an example by using a set of parameters: $\mu=84\text{ meV},\Delta_0=1\text{ meV},\alpha=25\text{ meV}$.
The energies of the first excited in-gap quasi-particle $E_{1}^{\pm}$ for different radius $R$ are summarized in Table.\ref{tab-finite-size-degeneracy-lift}.
For sufficient large radius $R=50\xi_0$, these two states shown in Eq.\eqref{eq-degenracy-state} is almost degenerated.
However, for smaller radius $R=30\xi_0$, the degeneracy of these two states will be totally lifted.

\begin{table}[!htbp]
\begin{center}
\begin{tabular}{ >{\centering\arraybackslash}m{0.4in}  >{\centering\arraybackslash}m{.85in} >{\centering\arraybackslash}m{.85in} >{\centering\arraybackslash}m{.85in}}
\toprule[1.5pt]
          & $\mathbf{R=50\xi_0}$ & $\mathbf{R=40\xi_0}$ & $\mathbf{R=30\xi_0}$ \\ \hline
\midrule
$E_1^{-}$ &  0.065047     & 0.053      &  0.038    \\
$E_1^{+}$ &  0.065062     & 0.078      &  0.093    \\
\bottomrule[1.5pt]
\end {tabular}
\caption{\label{tab-finite-size-degeneracy-lift} Energy for the first excited in-gap quasi-particle state $E_{1}^{\pm}$ for different radius. Parameters are: $\mu=84\text{ meV},\Delta_0=1\text{ meV},\alpha=25\text{ meV}$.}
\end{center}
\end{table}

\begin{figure*}[!htbp]
  \centering
  \includegraphics[width=6.0in]{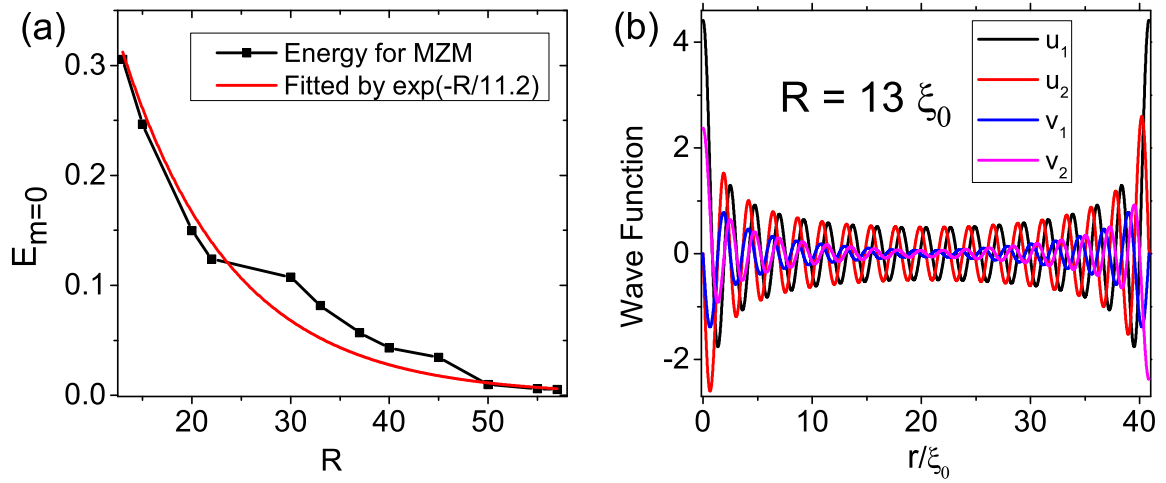}
  \caption{\label{fig-2-finite-size-effect} Two-vortex hybridization effect. (a) The energy of the $m=0$ state (MZM) for different sphere radius $R$. (b) The wave function of MZM for a small size $R=13\xi_0$.}
\end{figure*}

We now discuss the finite size effect on the two MZM state with tiny hybridization. The parameters mentioned in Sec.\ref{sec-2A-dispersion-wave} have been used to generate MZM, and the radius of sphere is $R=50\xi_0$,
which is large enough to separate the MZM in north pole and that in south pole. In Fig.~\ref{Fig-1-ADD-B-wave-func-MZM}, we show the wavefunction around both north pole and south pole as a function of $r=R\theta$,
which will decay to zero before reaching to equator of sphere (not shown in the figure). Recall that we use vortex-antivortex pair on sphere, i.e., $\Delta(\theta)=\Delta_0\tanh(R\sin\theta/\xi_0)$.
Now, if we change the distance between north pole (vortex) and south pole (anti-vortex) closer and closer by decreasing the radius $R$ of the sphere,
then we find in our calculation that the energy for MZM $E_0$ increases drastically to finite energy (order of superconductor gap)\cite{meng_prl_2009,sau_prb_2010,zhou_epl_2013}, shown in Fig.~\ref{fig-2-finite-size-effect}(a).

From Fig.~\ref{fig-2-finite-size-effect}(b), we can see that the wavefunction of MZM in north pole is hybridized with MZM in south pole.
As a result, the condition for MZM $\gamma=\gamma^\dagger$ is not satisfied in small radius situation (here we set $R=13\xi_0$).
And the two MZMs meet each other and are evolved into a normal electron or hole (say, complex fermions) with finite sub-energy gap.
For example, the finite energy $E_0$ is about $\pm0.3\Delta_0$ for $R=13\xi_0$ case. In this case, the MZM is not localized but looks like bulk state as shown in Fig.~\ref{fig-2-finite-size-effect}(b).
The loss of self-conjugate condition, together with the loss of localization condition, will give extremely different Andreev Reflection result, which will be discussed in Sec.\ref{sec-3C-sesar}.

As for the experiment data in Ref.\cite{xu_prl_2015}, it is found that a small external magnetic field (up to 0.18 Tesla) perpendicular to the surface will make the Abrikosov vortices closer and closer,
leading to weaker and weaker zero bias peak in STM/STS experiments. This can be easily understood from our model calculation,
because the hybridization of MZM's wave functions between two adjacent vortices will open a finite gap, see Fig.~\ref{fig-2-finite-size-effect}(a),
where black line is from our numerical calculation, and red line is fitted by function $\exp(-R/11.2)$.

\subsection{Local Density of States and Normal Conductance}\label{sec-2C-LDOS}
To analyze the experiment data from spin resolved STM, we should consider both normal conductance and Andreev reflection\cite{meir_prl_1992,sun_prb_1999,yang_prl_2010,huang_epl_2012}.
In this subsection, we shall calculate the local density of states (LDOS) and then estimate the normal conductance. As we know, the normal conductance is proportional to the local density of states $\mathcal{N}(E,r)$,
  \begin{align}
     \sigma_n(E,r) \equiv dI/dV(E,r) = \bar{\alpha} \mathcal{N}(E,r),
  \end{align}
where $\bar{\alpha}$ is assumed to be a constant. Under this approximation, we may estimate the normal conductance as follows,
\begin{align}\label{eq-normal-cond-estimation}
     \sigma_n(E<\Delta_0,r) = \sigma_n(\bar{E}\gg\Delta_0,r) \times \frac{\mathcal{N}(E<\Delta_0,r)}{\mathcal{N}(\bar{E}\gg\Delta_0,r)}.
\end{align}
\begin{figure}[!htbp]
  \centering
  \includegraphics[width=3.3in]{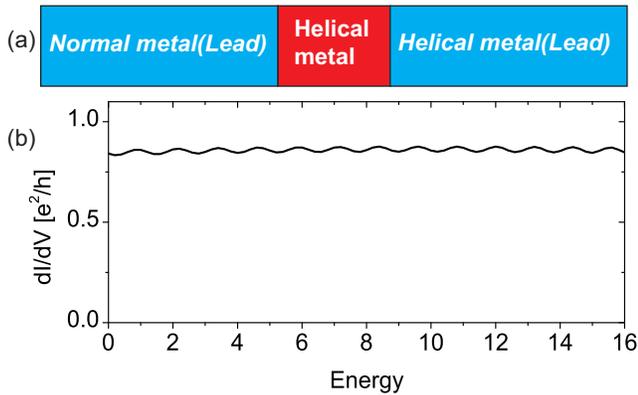}
  \caption{\label{fig-add-2-normal} (a) The normal metal(lead)/helical metal/helical metal(lead) junction used to estimate the normal conductance in STM/STS experiments.
  (b) Calculated normal conductance, $\sigma_n(\bar{E}\gg\Delta_0)\approx 0.88 \; e^2/h$, where we choose the hopping integral in the normal lead $t'=24$ and the coupling between the helical metal and left and right leads as $t_L=t'$ and $t_R=0.85t'$.}
\end{figure}

Here $\sigma_n(\bar{E}\gg\Delta_0)$ is the single particle tunneling conductance for the normal state. Since the superconducting gap is induced by the proximity effect ($\Delta_0$ is only 1 meV),
we may treat the normal state as helical metal when $\bar{E}\gg\Delta_0$ and $\epsilon(\mathbf{k})=\pm\sqrt{\left(\alpha\vert\mathbf{k}\vert\pm\mu\right)^2+\Delta_0^2}\approx\pm \left(\alpha\vert\mathbf{k}\vert\pm\mu\right)$ as in Ref.\cite{fu_prl_2008}.
Then we can set up a junction consisting of 1D normal lead/1D helical metal/1D helical metal lead (see Fig.~\ref{fig-add-2-normal}(a)) to calculate the single particle tunneling conductance for the helical metal.
By using recursive Green's function method and Landauer-B$\ddot{\text{u}}$tikker formula\cite{datta_book_1997},
we obtain $\sigma_n(\bar{E}\gg\Delta_0)\approx 0.88 \; e^2/h$ as shown in Fig.~\ref{fig-add-2-normal}(b).
Therefore, the normal conductance is given by
\begin{align}\label{eq:normal-cond}
     \sigma_n(E<\Delta_0,r) = \frac{\mathcal{N}(E<\Delta_0,r)}{\mathcal{N}(\bar{E}\gg\Delta_0,r)}\times 0.88 \; e^2/h.
\end{align}

\begin{figure*}[!htbp]
  \centering
  \includegraphics[width=5.8in]{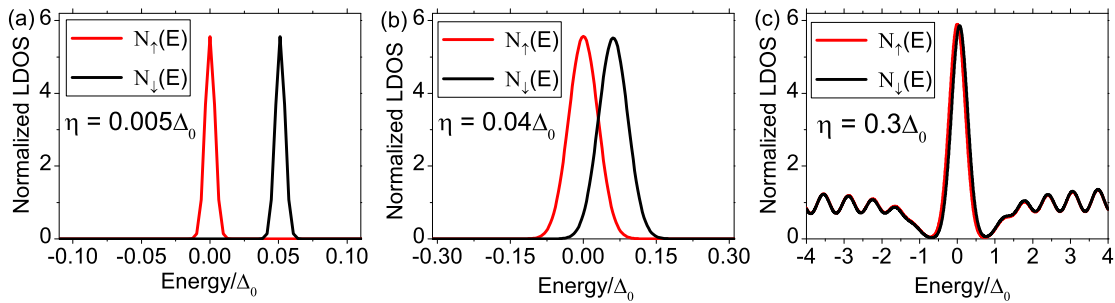}
   \caption{\label{fig-3-ldos}Local Density of States. (a) The smearing factor defined in Eq.\eqref{eq-delta-func-smearing} $\eta=0.005\Delta_0$. (b) $\eta=0.04\Delta_0\sim E_1$, comparable to the first excited state energy.
   (c) $\eta=0.3\Delta\approx 6E_1$. The LDOS for spin-up $\mathcal{N}_{\uparrow}(E)$ and spin-down $\mathcal{N}_{\downarrow}(E)$ coincide to each other.}
\end{figure*}

Thus the problem is reduced to calculate the spin polarized LDOS for quasi-particle excitations\cite{kawakami_prl_2015}, which is given by the electron wavefunctions,
  \begin{align}\label{eq:ldos-spin-up-and-down}
     \mathcal{N}(E,r)   &= \mathcal{N}_{\uparrow}(E,r) + \mathcal{N}_{\downarrow}(E,r), \\
     \mathcal{N}_{\uparrow}(E,r) &= \sum_{E_m} \left\lbrack \vert u_1\vert^2\delta(E-E_m) + \vert v_2\vert^2\delta(E+E_m)  \right\rbrack, \\
     \mathcal{N}_{\downarrow}(E,r) &= \sum_{E_m} \left\lbrack \vert u_2\vert^2\delta(E-E_m) + \vert v_1\vert^2\delta(E+E_m)  \right\rbrack,
  \end{align}
where $\delta(E-E_m)$ will be replaced by a smearing Gaussian function,
\begin{align}\label{eq-delta-func-smearing}
\delta(E-E_m)=\exp\left\{-(E-E_m)^2/\eta^2\right\}/(\sqrt{\pi}\eta),
\end{align}
where the smearing factor $\eta$ can be chosen smaller than the excitation gap $\Delta E=E_1-E_0$ in vortex core, as well as larger than $\Delta E$, which is the case in STM/STS experiments.

It is well known that the spectra of LDOS in vortex are discretized inside the gap and consist of several isolated peaks\cite{gygi_prb_1991,hayashi_prl_1998}.
Especially, there are only two bound-state peaks contributing to the spectra of LDOS at the vortex core center $r=0$.
Since LDOS is proportional to the amplitude square of wavefunction, and we have proved in Sec.~\ref{sec-2A-dispersion-wave}
that $\Phi_m({\theta=0})=\Phi_m({\theta=\pi})=0$ for $|m|>1$ (also see Fig.~\ref{Fig-1-ADD-B-wave-func-MZM} and Fig.~\ref{Fig-1-ADD-C-wave-func-the-first-excited-s}),
only the MZM ($E_0$) and the first excited state ($E_1$) will contribute to the bound-state peaks.

In Fig.~\ref{fig-3-ldos}, we plot the LDOS at $r=0$ for three values of the smearing factor $\eta$ in Eq.\eqref{eq-delta-func-smearing}.
For $\eta\ll E_1$, the LDOS for the spin-up ($\mathcal{N}_{\uparrow}$, MZM contribution) and for the spin-down ($\mathcal{N}_{\downarrow}$, the first excited state) are well separated in energy as shown in Fig.~\ref{fig-3-ldos}(a).
This also indicates the particle-hole asymmetry in vortex bound states as discussed in Ref.\cite{hayashi_prl_1998,li_sci-rep_2014,kawakami_prl_2015,mao_prb_2010}.
For $\eta\sim E_1$, $\mathcal{N}_{\uparrow}(E)$ and $\mathcal{N}_{\downarrow}(E)$ overlap but are still distinguishable as plotted in Fig.~\ref{fig-3-ldos}(b).
For $\eta\gg E_1$, $\mathcal{N}_{\uparrow}(E)$ and $\mathcal{N}_{\downarrow}(E)$  become essentially the same. Note that $\eta$ is a measure of the energy resolution in the STM experiment,
which is presently poor to distinguish $E_1$ from the zero mode. We believe that Fig.~\ref{fig-3-ldos}(c) corresponds to the experiment situation in Ref.\cite{arxiv_mzm_our_2016}.
It is a very important conclusion in this article. We also note that the similar result was reported in Ref.\cite{kawakami_prl_2015}.

\section{Andreev reflection}\label{sec-3-AR}
In this section, we present the method and results for the calculation of Andreev Reflection(AR) based on the solution of BdG equations, and explain recent experiments in Ref.\cite{arxiv_mzm_our_2016}.
We consider the tiny STM tip as a 1D metallic lead and ``touch" the vortex core as a single point contact.
We argue that this should be good approximation in the high barrier limit. The s-wave superconducting gap will induce effective triplet pairing correlation in the bulk spectrum for the system with strong spin-orbit coupling\cite{alicea_prb_2010,chiu_arxiv_2016}.
However, it will not affect our discussions on spin selective Andreev reflection (SSAR) for MZM, because we only focus on the center of the the vortex core, where the superconducting gap is zero.

Majorana zero mode (MZM)\cite{vic_prl_2009,lutchyn_prl_2010} will contribute to the measurement of $dI/dV$ for 1D nanowire system via Andreev reflection\cite{wimmer_njp_2011}.
Moreover, He \emph{et al.} figured out that spin selective Andreev reflection (SSAR) can be used to reveal spin degree of freedom for MZM, and they demonstrated it in a 1D nanowire system \cite{he_prl_2014}.
However, there are two disadvantages to probe SSAR in 1D nanowire in experiments.
Firstly, the signal for Andreev reflection of MZM would be mixed with the normal Andreev reflection of s-wave superconductor.
Secondly, it requires strong Zeeman splitting ($V_z^2>\mu^2+\Delta^2$) in the system\cite{alicea_rep_pro_2012}, which may affect the spin property of MZM and suppress the superconductivity.
In this paper, we generalize He \emph{et al.}'s theory to TI/SC heterostructure modelled in Eq.~\eqref{eq:bdg-ham-ti-sc}, and find similar SSAR inside the vortex core.

\subsection{Transport method for a Normal metal/Superconductor junction}\label{sec-3A-transport-method}

To study SSAR contribution to total conductance measured in spin polarized STM/STS experiments, we utilize the setup sketched in Fig.~\ref{fig-add-1-sketch-sphere} and Fig.~\ref{Fig-ADD-sketch-sesar}.
Since the STM tip size in STM/STS experiments is about 0.01 nm, which is much smaller than the vortex size $\xi_0=35$ nm, we can treat the STM tip as a normal lead and the contact between STM tip and TI/SC as point contact.
The normal lead can be described by the following Hamiltonian,
  \begin{align}
     \mathcal{H}_{L} = &\sum_{i<0,\sigma} \left\{ -t' \hat{c}^\dagger_{Li\sigma}\hat{c}_{Li+1\sigma}  + \text{H.c.}\right\} \\ \nonumber
     &+ \left\{ (2t' - \mu')\hat{c}^\dagger_{Li\sigma}\hat{c}_{Li\sigma}+ \hat{c}^\dagger_{Li\alpha} \lbrack\vec{V}\cdot\vec{\sigma}\rbrack_{\alpha,\beta}\hat{c}_{Li\beta}\right\}.
  \end{align}
The coupling between the normal lead and TI/SC (the contact point locates inside the vortex core) is given by
  \begin{align}\label{eq-ham-coupling-lead-device}
     \mathcal{H}_t = \sum_{\sigma}\left\{ t_c\hat{c}^\dagger_{L0\sigma}\hat{c}_{S1\sigma} +  \text{H.c.} \right\}.
  \end{align}

To proceed, we shall demonstrate the SSAR induced by MZM in the high barrier limit and choose the parameters as follows, $\vert\vec{V}\vert = 10\Delta_0$, $t'=25\Delta_0$, $\mu'=0$ and $t_c=0.008\Delta_0$.
Note that the coupling strength $t_c$ only affects the width of zero bias peak of MZM, which is of the Lorentz shape\cite{he_prl_2014}, $dI/dV \sim E^2/(E^2+\tilde{t}^2)$, where $\tilde{t}\sim t_c$.
Other parameters in the TI/SC heterostructure are chosen as the same as those in Sec.\ref{sec-2A-dispersion-wave}.

With the help of the solution to Eq.~\eqref{eq:eigenvalues-problem}, we are able to calculate the retarded Green's function for the single-particle system,
\begin{align}\label{eq-green-function-definition}
      G^{R}_{0} (E, \vec{r}, \vec{r}') = \sum_{m}\sum_{n} \frac{\vert \Phi_m^{n} (\vec{r})\rangle\langle\Phi_m^{n}(\vec{r}')\vert}{E-E_m^{n}+i\delta},
\end{align}
where $m$ is the angular momentum (quantum number of $K_z$ in Eq.~\eqref{eq-total-angular-kz}), $n$ is an additional quantum number labelling the eigen-states,
$\delta$ is positive infinitesimal and set as $10^{-5}$meV in our calculation, and $G^{R}_{0}$ is a $4\times 4$ matrix.

For the point contact problem, we need to evaluate the local Green's function  $G^{\text{tot}}(E,\vec{r},\vec{r})$ in the coupled system.
Considering the $\delta$-function interaction between normal lead and the 2D TI/SC heterostructure, we can write $G^{\text{tot}}(E,\vec{r},\vec{r})$ in terms of $G^{R}_0(E,\vec{r},\vec{r})$ and self-energy $\Sigma(E,\vec{r})$ through the Dyson equation\cite{dyson_phys-rev_1949,datta_book_1997},
   \begin{align}\label{eq:total-green-function}
      G^{\text{tot}}(E,\vec{r},\vec{r}) = \frac{1}{\left(G^{R}_0(E,\vec{r},\vec{r})\right)^{-1}-\Sigma(E,\vec{r})},
   \end{align}
or its iterative form
   \begin{align}
     G^{\text{tot}}(E,\vec{r},\vec{r}) = G^{R}_0(E,\vec{r},\vec{r}) + \Sigma(E,\vec{r}) G^{\text{tot}}(E,\vec{r},\vec{r}).
   \end{align}
After that, the $S$-matrix for the junction can be calculated by Fisher-Lee formula\cite{landauer_mag_1970,fish-lee_prb_1981,buttiker_prb_1988},
   \begin{align}\label{eq:$S$-matrix}
      \tilde{S} = -\text{I} + i\Gamma^{1/2} \times G^{\text{tot}} \times \Gamma^{1/2},
   \end{align}
where the broadening function $\Gamma$ is defined as $\Gamma = i (\Sigma - \Sigma^{\dagger})$, which is a $4\times 4$ matrix too. We can read out the $2\times 2$ reflection matrices \(\tilde{r}_{ee}\) and \(\tilde{r}_{he}\) from the $S$-matrix,
\begin{align}
\tilde{S} = \left( \begin{array}{cc} \tilde{r}_{ee} & \tilde{r}_{eh} \\ \tilde{r}_{he} & \tilde{r}_{hh} \end{array} \right),
\end{align}
where $\tilde{r}_{ee}^{\sigma,\sigma'}$ means that a spin-$\sigma'$ electron incomes and a spin-$\sigma$ electron outgoes.
Thus, we can calculate the differential conductance coming from Andreev reflection $dI/dV$ using Landauer-B$\ddot{\text{u}}$tikker formula\cite{datta_book_1997},
\begin{align}\label{eq:andreev-cond}
        \sigma_A(E,\vec{r})\equiv dI/dV(E,\vec{r}) &=  \text{Tr}\left\lbrack \tilde{r}_{he}^\dagger \tilde{r}_{he} \right\rbrack \times 2e^2/h.
\end{align}

\subsection{Spin selective Andreev reflection}\label{sec-3C-sesar}
In this subsection, we will discuss the SSAR effect and focus on the vortex core center $r=0$, where the MZM spin is parallel to the magnetic field (spin-up) by symmetry.
The case of $r=0$ also means that we only need to keep $m=0$ channel in Eq.~\eqref{eq-green-function-definition}, which will greatly simplify our $dI/dV$ calculations.

\begin{figure}[!htbp]
  \centering
  \includegraphics[width=3.1in]{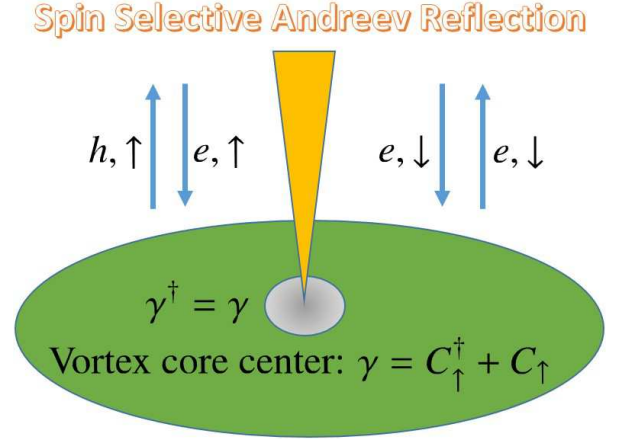}
  \caption{\label{Fig-ADD-sketch-sesar} Illustration of spin selective Andreev reflection. an incoming spin-down electron will be reflected as spin-down electron because of the mismatch of the spins of the incoming electron and the MZM. }
\end{figure}

As pointed out by He \emph{et al.}\cite{he_prl_2014}, an incoming spin-up electron will be reflected as a spin-up hole,
while an incoming spin-down electron will be reflected as spin-down electron because of the mismatch of the spins of the incoming electron and the MZM.
This phenomenon is called spin selective Andreev reflection (SSAR)\cite{he_prl_2014,arxiv_mzm_our_2016}, as illustrated in Fig.~\ref{Fig-ADD-sketch-sesar}.
Note that the superconducting gap vanishes at the vortex core center $r=0$, therefore the Andreev reflection is via the MZM only.

We would like to emphasize that the existence of a localized MZM is the necessary condition for observing SSAR in experiments.
So that SSAR serves a definite experimental evidence for MZM. To examine this point, we shall study SSAR in our model with two different sphere radius sizes, $R=50\xi_0$ and $R=13\xi_0$.
As discussed in Sec.~\ref{sec-2B-finite-size-effect}, (i) when $R=50\xi_0$, there exists a single MZM inside each vortex and anti-vortex core, which is well separated from each other;
(ii) when $R=13\xi_0$, two MZMs will hybridize with each other strongly, resulting in usual complex fermions with extended spatial wavefunctions.

\begin{figure}[!htbp]
  \centering
  \includegraphics[width=3.3in]{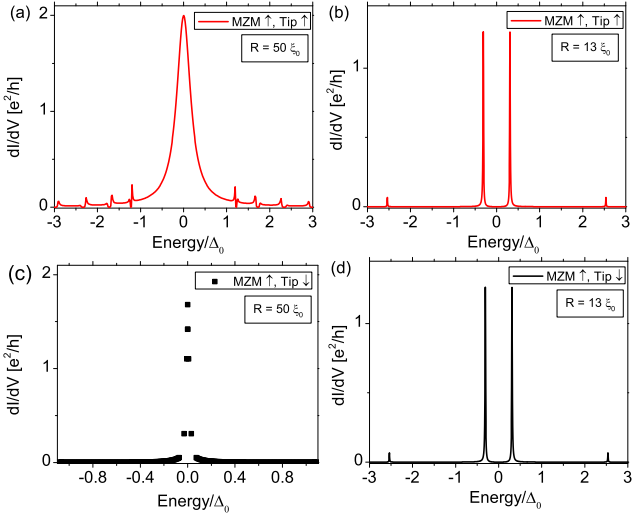}
  \caption{\label{fig-4-andreev-reflection} Calculated AR conductance $dI/dV$ at vortex core center $r=0$.
  (a) $R=50\xi_0$, the incoming electron spin polarization is parallel to local MZM spin.
  (b) $R=13\xi_0$£¬the incoming electron spin polarization is parallel to local MZM spin.
  (c) $R=50\xi_0$, the incoming electron spin polarization is antiparallel to local MZM spin.
  (d) $R=13\xi_0$£¬the incoming electron spin polarization is antiparallel to local MZM spin.
  }
\end{figure}

Numerical results for AR conductance $dI/dV$ are plotted in Fig.~\ref{fig-4-andreev-reflection}.
For the large radius $R=50\xi_0$ with a well localized MZM, when the incoming electron spin polarization is parallel to the MZM spin at vortex core center,
AR conductance $dI/dV$ exhibits a zero bias peak with quantized conductance $2e^2/h$ and significant weight in the spectra, as shown in Fig.~\ref{fig-4-andreev-reflection}(a).
However, for the small radius $R=13\xi_0$ with two strongly hybridized MZMs, $dI/dV$ only have two sharp peaks with vanishing weight as plotted in Fig.~\ref{fig-4-andreev-reflection}(b).
These artificial peaks correspond to two hybridized MZMs numerically, and can not be observed in experiments even at very low temperatures.
Indeed, similar sharp peaks also appear in the ordinary Andreev reflection in $s$-wave superconductors at superconducting gap edge in the high barrier limit, see Appendix \ref{sec-3B-BTK-theory} for details.
On the contrary, when the incoming electron spin polarization is antiparallel to the MZM spin at vortex core center, the signal of AR conductance is completely suppressed for both large and small radius, as shown in Fig.~\ref{fig-4-andreev-reflection}(c) and (d).
From these results, one can conclude that the Andreev reflection is spin selective at vortex core center in the presence of a localized MZM, while AR conductance is completed suppressed at the vortex core center when the MZM vanishes due to hybridization or other reasons.

It is noted that the results in Fig.~\ref{fig-4-andreev-reflection} also explain the experimental fact that all the zero bias peaks in $dI/dV$ of AR will disappear when the external magnetic field exceeds a small threshold value.
The reason is the following. Increasing the external magnetic field will shorten the distance between two neighboring vortices on the Abrikosov lattice. Then a MZM inside a vortex core will hybridize with another on in the neighboring vortex core, leading to complex fermions instead of MZMs.

\begin{figure*}[hptb]
  \centering
  \includegraphics[width=6.0in]{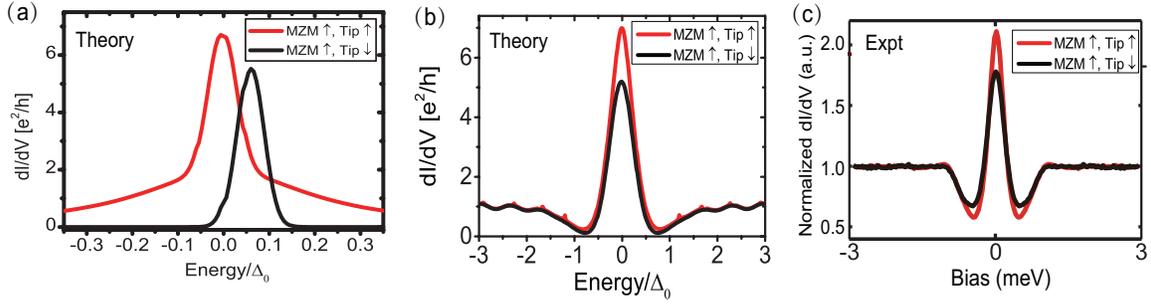}
  \caption{\label{fig-6-total-cond} (a) Calculated total conductance with $\eta=0.04\Delta_0$. (b) Calculated total conductance with $\eta=0.3\Delta_0$. (c) Experiment data from Ref.\cite{arxiv_mzm_our_2016}.}
\end{figure*}

\subsection{Total differential conductance in spin polarized STM/STS experiments}\label{sec-3D-total-cond}

The total conductance in spin polarized STM/STS experiments can be evaluated by adding the normal component $\sigma_n(E,r)$ in Eq.\eqref{eq:normal-cond} to the AR component $\sigma_A(E,r)$ in Eq.\eqref{eq:andreev-cond},
   \begin{align}\label{eq-cond-total}
      \sigma_{\text{tot}}(E,r) = \sigma_n(E,r) + \sigma_A(E,r).
   \end{align}
Although the STM tip can be located in arbitrary $r$ in STM/STS experiments, the SSAR effect will be suppressed and then vanish as $r$ increases, when the differential conductance $dI/dV$ is no longer spin dependent.
So that let us focus on $r=0$ at first, and then discuss how $r>0$ will change the results.

The numerical results at $r=0$ are shown and compared with experimental data in Fig.~\ref{fig-6-total-cond}.
In our calculations, we choose two different energy smearing factors defined in \eqref{eq-delta-func-smearing}, $\eta=0.04\Delta_0$ and $\eta=0.3\Delta_0$.
The former is chosen as close to the first in-gap excited energy $E_1$, while the latter is close to energy resolution in STM/STS experiments\cite{arxiv_mzm_our_2016}.
The calculated total conductance $dI/dV$ for $\eta=0.04\Delta_0$ and $\eta=0.3\Delta_0$  is plotted in Fig.~\ref{fig-6-total-cond}(a) and Fig.~\ref{fig-6-total-cond}(b) respectively,
and the experimental result\cite{arxiv_mzm_our_2016} is plotted in Fig.~\ref{fig-6-total-cond}(c) for comparison.

For $\eta=0.04\Delta_0\sim E_1$, the total conductance $dI/dV$ exhibits two distinguishable peaks for different spin polarization, see Fig.~\ref{fig-6-total-cond}(a).
When the STM tip spin polarization is parallel to the MZM spin (is parallel to external magnetic field at $r=0$), which is denoted as ``MZM $\uparrow$, Tip $\uparrow$'', there is a zero bias peak.
When the tip spin polarization is antiparallel to the MZM spin, denoted as ``MZM $\uparrow$, Tip $\downarrow$'', there is a peak around $E_1$.
Hereafter we shall denote the height of these two peaks as $dI/dV\vert_\uparrow$ and $dI/dV\vert_\downarrow$ respectively.
The separation of these two peaks in energy is due to the LDOS, see Fig.~\ref{fig-3-ldos}, which contributes to the normal conductance $\sigma_n$.

For $\eta=0.4\Delta_0\gg E1$, which is close to the experimental situation, there exist two zero bias peak in $dI/dV$ as shown in Fig.~\ref{fig-6-total-cond}(b), with height $dI/dV\vert_\uparrow$ and $dI/dV\vert_\downarrow$.
Since LDOS is almost spin independent in this case, $\mathcal{N}_\uparrow=\mathcal{N}_\downarrow$, therefor the normal conductance is spin independent too and the difference between $dI/dV\vert_\uparrow$ and $dI/dV\vert_\downarrow$
comes from SSAR entirely. The spin polarization of tunneling conductance is estimated as
\begin{align}
   P = \frac{dI/dV\vert_\uparrow - dI/dV\vert_\downarrow}{dI/dV\vert_\uparrow + dI/dV\vert_\downarrow} \sim 16\%,
\end{align}
which is about 2.3 times of the experimental value $7\%$\cite{arxiv_mzm_our_2016}, see Fig.~\ref{fig-6-total-cond}(c) for details.
The deviation from experimental value may be due to disorder effect, which has not yet considered in this paper.

\begin{figure}[!htbp]
  \centering
  \includegraphics[width=3.1in]{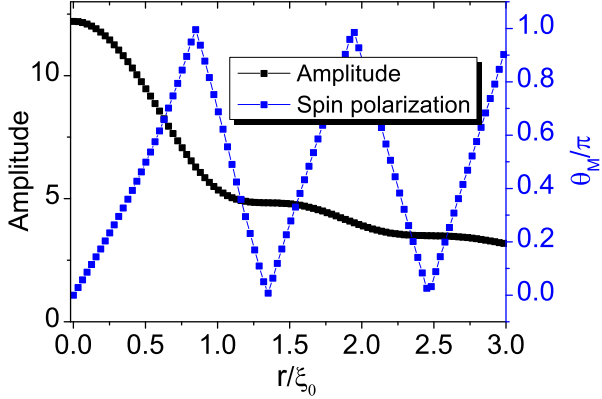}
  \caption{\label{fig-6-total-didv-spin-direction} The angle between the local MZM spin and external magnetic field $\theta_M$ and the amplitude of MZM wavefunction $\sqrt{\vert u_1(r)\vert^2+\vert u_2(r)\vert^2}$.
  $\theta_M=0$ is for parallel and $\theta_M=\pi$ is for antiparallel.}
\end{figure}

Now let us discuss the situation when $r>0$. Due to spin-orbit coupling, spin is not a good quantum number of MZM and will vary spatially. On the other hand, the amplitude of the MZM wavefunction, $\vert u_1\vert^2 + \vert u_2\vert^2$, will decay as $r$ increase.
The angle between the local spin direction of MZM and external magnetic field $\theta_M$ is plotted as a function of $r$ as well as the amplitude $\vert u_1\vert^2 + \vert u_2\vert^2$ in Fig.~\ref{fig-6-total-didv-spin-direction}.
When the STM tip moves away from the vortex core center, two reasons will reduce the SSAR signal. Firstly, the amplitude of MZM wavefunction becomes smaller and smaller. Secondly, the mismatch between STM tip spin polarization and the local spin of MZM will reduce
the AR conductance via MZM. This explains the experimental observation that the spin dependence of $dI/dV$ becomes too weak to detect at about $r=0.3\xi_0$\cite{arxiv_mzm_our_2016}.

\section{Conclusions} \label{sec-4-conclusion}
In this article, we use the proposal by Fu and Kane\cite{fu_prl_2008} to generate the MZM in the vortex core on the interface between TI and SC. Then, we take the parameters derived from the experiment to solve the BdG equation in Eq.\eqref{eq:bdg-ham-ti-sc}. Based on the results for both larger radius and smaller radius, we simulate the distance between the vortex in north pole and the vortex in south pole, so that we can discuss the two-vortices problem(finite size effect). It may be related to the experiment, in which the external magnetic field will make the vortices closer and closer so that the ZBP will disappear. We think it is because of the hybridization of MZMs' wave functions in nearby vortices, which will lead to a finite sub-energy gap. In addition, we calculate the LDOS for small smearing factor and larger smearing. The asymmetry of LDOS has not be seen in experiment so far, because the STM energy resolution(0.1 meV) is larger than the minigap(0.05 meV). Finally, we use Green function's approach to calculate the $S$-matrix for N/S junction by Fish-Lee-Landauer-B$\ddot{\text{u}}$tikker formula, and we find similar SSAR effect in our model calculation. Finally, we also estimate the total conductance $dI/dV$ in our calculation is qualitatively consistent with the experiment\cite{arxiv_mzm_our_2016}.

Furthermore, we wish to point out that the estimation for normal conductance here is considered in an approximation way. Precise calculation may change our results a little, but the main physics should be kept as we discussed in this article, due to the spin property of MZMs.

\section{Acknowledgement}
We acknowledge helpful discussions with Chih-Chieh Chen, Chui-Zhen Chen, Fei Ye, Jin-Hua Gao, Wei-Qiang Chen and K.T. Law. Especially, we benefit a lot from detailed communications with James Jun He.
This work is supported in part by National Basic Research Program of China (No.2014CB921201/2014CB921203), National Key R\&D Program of the MOST of China (No.2016YFA0300202),
NSFC (No.11374256/11274269) and the Fundamental Research Funds for the Central Universities in China. F.C.Z was also supported by the Hong Kong¡¯s University Grant Council via Grant No. AoE/P-04/08.

\begin{appendix}
\section{Reproduce BTK theory}\label{sec-3B-BTK-theory}
In this appendix, we shall discuss the vortex-free case in details. In this case, the superconducting gap is a constant, i.e., $\Delta(\theta,\phi) = \Delta_0$.
Although the analytical solution to Hamiltonian Eq.~\eqref{eq-bdg-ho-hdelta-vortex-free-1} has been derived in Eq.~\eqref{eq-solution-vortex-free-energy},
we would like to use the numerical method, discussed in the main text in Sec.\ref{sec-1-models-methods}, to solve this problem once more and reproduce the results of BTK theory for double check.
For a conventional $s$-wave superconductor, the kinetic term may involve $\eta L^2/R^2$ and drop out linear term. The Hamiltonian reads,
\begin{align}
  \mathcal{H}_{\text{BdG}}
  &= \left(\frac{\eta}{R^2}L^2-\frac{\alpha}{R}\vec{L}\cdot\vec{\sigma} -\mu\right)\otimes\tau_z + \Delta_0 I\otimes\tau_x,
\end{align}
which can be written in the form of a $4\times 4$ matrix,
\begin{widetext}
\begin{align}\label{eq-ham-vortex-free}
                    \mathcal{H}_{\text{BdG}}=\left(
                      \begin{array}{cccc}
                        \frac{\eta}{R^2}L^2-\mu-\frac{\alpha}{R}L_z  & -\frac{\alpha}{R}L_- & \Delta_0 & 0 \\
                        -\frac{\alpha}{R}L_+ & \frac{\eta}{R^2}L^2-\mu+\frac{\alpha}{R}L_z  & 0 & \Delta_0 \\
                        \Delta_0 & 0 & -(\frac{\eta}{R^2}L^2-\mu)+\frac{\alpha}{R}L_z  & \frac{\alpha}{R}L_-  \\
                        0 & \Delta_0 & \frac{\alpha}{R}L_+  & -(\frac{\eta}{R^2}L^2-\mu) -\frac{\alpha}{R}L_z  \\
                      \end{array}.
                    \right)
\end{align}
\end{widetext}

\begin{figure}[!htbp]
  \centering
  \includegraphics[width=8cm]{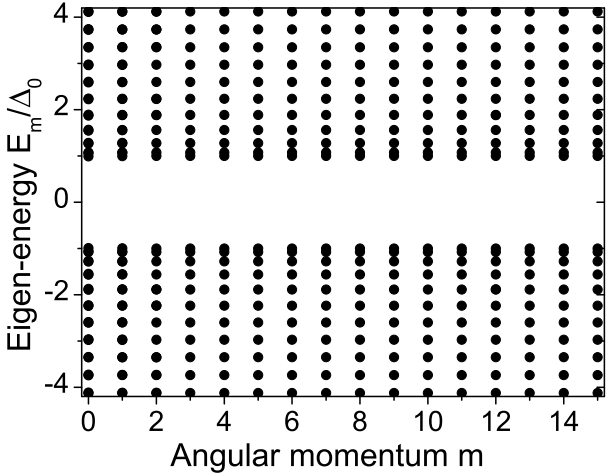}
  \caption{\label{fig-add-dispersion-vortex-free} Energy spectra for vortex-free case. Only $E_m$ for non-negative angular momentum are calculated numerically and plotted here, with used parameters $\Delta_0=1$ meV, $\eta=20$ meV, $\xi_0=35$ nm, $R=50\xi_0$, $\alpha=0.1$ meV and $\mu=32$ meV.}
\end{figure}

\begin{figure*}[!htbp]
  \centering
  \includegraphics[width=17cm]{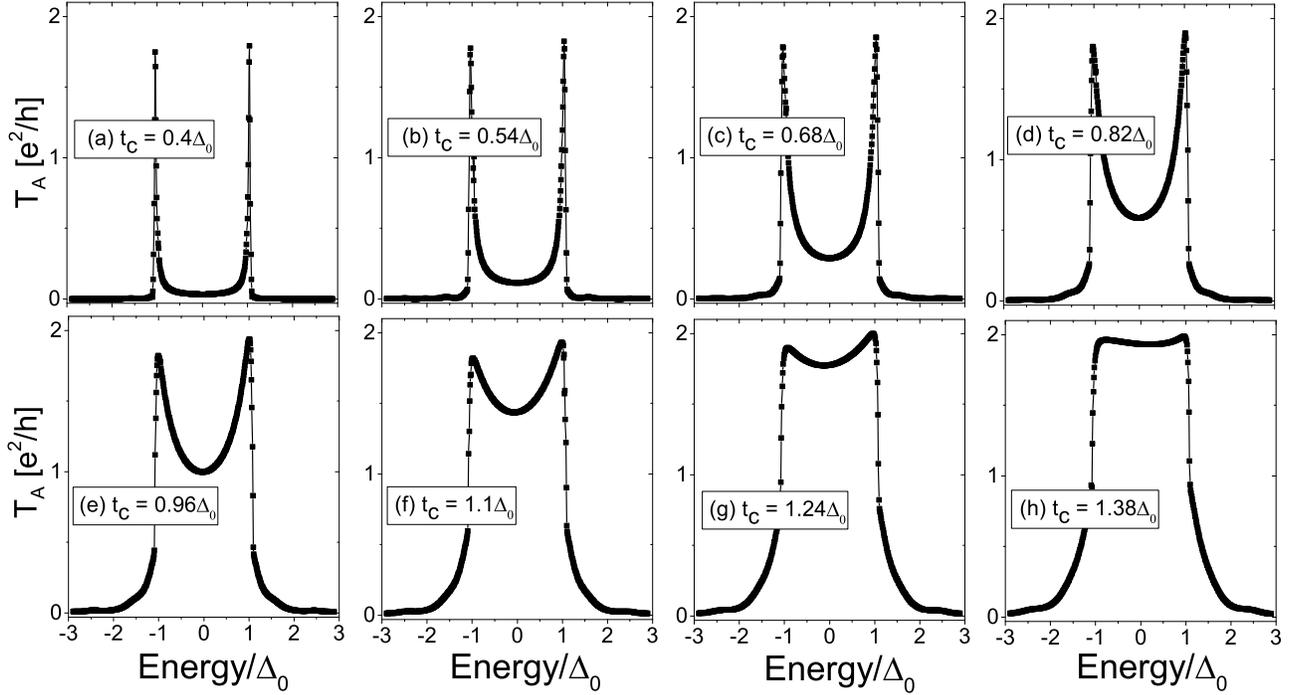}
  \caption{\label{fig-5-BTK} Ordinary Andreev reflection calculated using $T_A=\text{Tr}\left\lbrack \tilde{r}_{he}^\dagger \tilde{r}_{he} \right\rbrack$ to reproduce the BTK theory (two channels/particles scattering process). Increase the coupling $t_c$ from $0.4\Delta_0$ (high barrier case, large $Z$) in (a) to $t_c=1.38\Delta_0$(transparent limit, tiny $Z$) in (h).}
\end{figure*}

Since $J_z$ is still a good quantum number in this situation, the same technique to solve the BdG equation in the main text is still valid here,
expect that the basis in Eq.\eqref{eq:two_good_quamtum_operator_h_jz} and Eq.\eqref{eq:two_good_quamtum_operator_h_jz-2} will change. Now the secular equations become
\begin{align}\label{eq:good-quantum-vortex-free}
      \mathcal{H}  \left( \begin{array}{c}
                    e^{im\phi}\;u_1 \\
                    e^{i(m+1)\phi}\;u_2 \\
                    e^{im\phi}\;v_1 \\
                    e^{i(m+1)\phi}\;v_2
                  \end{array} \right) &= E \left( \begin{array}{c}
                    e^{im\phi}\;u_1 \\
                    e^{i(m+1)\phi}\;u_2 \\
                    e^{im\phi}\;v_1 \\
                    e^{i(m+1)\phi}\;v_2
                  \end{array} \right),   \\
       J_z \left( \begin{array}{c}
                    e^{im\phi}\;u_1 \\
                    e^{i(m+1)\phi}\;u_2 \\
                    e^{im\phi}\;v_1 \\
                    e^{i(m+1)\phi}\;v_2
                  \end{array} \right) &= m \left( \begin{array}{c}
                    e^{im\phi}\;u_1 \\
                    e^{i(m+1)\phi}\;u_2 \\
                    e^{im\phi}\;v_1 \\
                    e^{i(m+1)\phi}\;v_2
                  \end{array} \right),
\end{align}
where we still use the standard Nambu representation $\left(c_{\uparrow},c_{\downarrow},c_{\downarrow}^\dagger, -c_{\uparrow}^\dagger\right)$ as in the main text.
Denote \(\vec A=(a_{l_1},\cdots,a_{l_N})^{T}\;,\; \vec B=(b_{l_1},\cdots,b_{l_N})^{T} \;,\; \vec C=(c_{l_1},\cdots,c_{l_N})^{T}\;,\; \vec D=(d_{l_1},\cdots,d_{l_N})^{T} \), the eigenvalues problem can be expressed as follows,
\begin{align}
      \mathcal{H} \left( \begin{array}{c}
                    \vec A \\
                    \vec B \\
                    \vec C \\
                    \vec D
                  \end{array} \right) = E  \left( \begin{array}{c}
                    \vec A \\
                    \vec B \\
                    \vec C \\
                    \vec D
                  \end{array} \right).
\end{align}
Choosing the parameters as follows, $\Delta_0=1$ meV, $\eta=20$ meV, $\xi_0=35$ nm, $R=50\xi_0$, $\alpha=0.1$ meV and $\mu=32$ meV, and using similar cutoff as the vortex pair case, we are able to solve these equations numerically.
Note that the Rashba coupling $\alpha$ is sufficient small, therefore it will not change the $s$-wave superconducting pairing qualitatively.

The energy spectra are shown in Fig.\ref{fig-add-dispersion-vortex-free},
which is consistent with the analytical results for the superconducting gap in Eq.\eqref{eq-solution-vortex-free-energy}. It is easy to calculate the Green's function from obtained eigen-states, in order to reproduce the BTK theory.

Now we shall calculate the reflection coefficient for Andreev reflection.
To do this, we treat the STM tip as a normal lead, and the Hamiltonian $\mathcal{H}_{L}$ for this lead reads,
\begin{align}
     \mathcal{H}_{L} &= \sum_{i<0,\sigma}\Big{\{} \left\lbrack -t' \hat{c}^\dagger_{Li\sigma}\hat{c}_{Li+1\sigma} + \text{H.c.}\right\rbrack \nonumber \\
      &\qquad\quad + \left\lbrack (2t' - \mu')\hat{c}^\dagger_{Li\sigma}\hat{c}_{Li\sigma}\right\rbrack \Big{\}}.
\end{align}
The coupling between the superconductor (contact point locates in the vortex core center) and the lead Hamiltonian is given by,
\begin{align}
     \mathcal{H}_t = \sum_{\sigma}\left\{ t_c\hat{c}^\dagger_{L0\sigma}\hat{c}_{S1\sigma} +  \text{H.c.} \right\},
\end{align}
where we choose the parameters for the 1D normal Lead (STM tip) as: \(t'=25\Delta_0,\; \mu'=24\Delta_0\).
The coupling constant \(t_c\) can be tuned from transparent limit to high barrier limit.

Using the technique introduced in Sec.\ref{sec-3A-transport-method}, we calculate the Andreev reflection coefficient $T_A=\text{Tr}\left\lbrack \tilde{r}_{he}^\dagger \tilde{r}_{he} \right\rbrack$ for various coupling constants $t_c$.
The numerical results are shown in Fig.~\ref{fig-5-BTK}(a-h). The transparent limit will be taken when $t_c\ge \Delta_0$, while the high barrier limit occurs at $t_c\ll \Delta_0$.

Now we would like to compare our numerical results with BTK theory for N/I/S junction\cite{BTK_physics_1982,tinkham_book_2004,kashiwaya_rep_pro_2000}.
In the BTK theory, the Andreev reflection coefficient $T_A$ is evaluated through matching boundary condition,
\begin{align}
   T_A = \begin{cases}
      E < \Delta: \quad \frac{\Delta^2}{E^2+(\Delta^2-E^2)(1+2Z^2)^2}, \\
      E > \Delta: \quad \frac{u_0^2v_0^2}{\gamma^2},
   \end{cases}
\end{align}
where $\gamma^2 = u_0^2 + Z^2(u_0^2-v_0^2)$ and $u_0^2=1-v_0^2=\frac{1}{2}\left\lbrack 1+\sqrt{\frac{E^2-\Delta^2}{E^2}} \right\rbrack$,
$Z$ is the barrier strength and gives rise to the contact potential \(Z\delta(r)\) at the interface between normal metal and superconductor\cite{kashiwaya_rep_pro_2000}.
$Z=0$ corresponds to the transparent limit, resulting in complete reflection inside the superconducting gap, $T_A(E\le\Delta)=1$.
On the contrary, $Z\to \infty$ corresponds to the high barrier limit. In this limit, $T_A(E<\Delta)=0$ but $T_A(E=\Delta)=1$, namely, Andreev reflection only happens at the edge of the superconducting gap.
It is clear that our numerical results reproduce BTK theory well, except the maximum value of $T_A$ is 2 instead of 1. This is because we count two channels in our model.

\end{appendix}

\bibliography{reference}

\end{document}